\documentclass{aa}
\usepackage{graphicx,times}

\newcommand{\xmm}{{\it XMM-Newton}}
\newcommand{\chandra}{{\it Chandra}}
\newcommand{\rgs}{{RGS}}
\newcommand{\epic}{EPIC}
\newcommand{\mos}{EPIC-MOS}
\newcommand{\pn}{EPIC-PN}
\newcommand{\acis}{ACIS}
\newcommand{\n}{SNR N\,103B}

\begin{document}

\title{High-Resolution X-ray imaging and spectroscopy of \n}

\author{K.J. van der Heyden\inst{1},
        E. Behar\inst{2},
        J. Vink\inst{2,3},
        A.P. Rasmussen\inst{2},
        J.S. Kaastra\inst{1},
        J.A.M. Bleeker\inst{1},
        S.M. Kahn\inst{2},
        R. Mewe\inst{1}}
\offprints{K.J. van der Heyden}

\institute{SRON National Institute for Space Research, Sorbonnelaan 2, 
3584 CA Utrecht, The Netherlands 
\and 
Columbia Astrophysics Laboratory, Columbia University, 550 West 120th 
Street, 
New York, NY 10027, USA 
\and 
\chandra\ fellow 
\\email: K.J.van.der.Heyden@sron.nl ; behar@astro.columbia.edu ; jvink@astro.columbia.edu ; arasmus@astro.columbia.edu ; J.S.Kaastra@sron.nl ; J.A.M.Bleeker@sron.nl ; skahn@astro.columbia.edu ; R.Mewe@sron.nl \\} 
 
\titlerunning{\n} 
\authorrunning{K.J. van der Heyden et al.} 
 
\date{Received ; accepted } 
 
\abstract{ The X-ray emission from the young supernova remnant (SNR) \n\ is
measured and analysed using the high-resolution cameras and spectrometers on
board \xmm\ and \chandra. The spectrum from the entire remnant is reproduced
very well with three plasma components of $kT_{\rm e}$ = 0.55, 0.65, and
3.5~keV, corresponding roughly to line emission by the O-K, Fe-L, and Fe-K
species, respectively. Narrow band images reveal different morphologies for
each component. The $kT_{\rm e}$ = 0.65~keV component, which dominates the
emission measure (4.5$\times$10$^{65}$m$^{-3}$), is in ionisation equilibrium.
This provides a lower limit of 3000 yrs to the age of the remnant, which is
considerably older than the previously assumed age of the remnant (1500~yrs).
Based on the measured energy of the Fe-K feature at 6.5~keV, the hot (3.5~keV)
component is found to be recently shocked ($\sim$200~yrs) and still ionising.
The high elemental abundances of O and Ne and the low abundance of Fe could
imply that \n\ originated from a type II supernova (SN) rather than a type Ia SN as previously thought. 
 }

\maketitle

\section{Introduction} 
 
The young supernova remnant (SNR) N\,103B (also known as SNR 0509-69) is the
fourth brightest X-ray remnant in the Large Magellanic Cloud (LMC). It was
first identified as a supernova remnant by Mathewson and Clarke
(\cite{mathewson}) based on the non-thermal nature of its radio spectrum and the
relative emission-line strengths of \ion{S}{ii} and H$\alpha$. Due to its small
size (3~pc), Hughes et al.~(\cite{hughes}) estimated the age of \n\ to be about
1500~yrs. Recent \chandra\ data (Lewis et al.~\cite{lewisa}) show that the
remnant is much brighter on the western limb than towards the East, this is most
likely because of density contrasts between East and West. The radio and X-ray
brightness variations are correlated on large scales, but do not overlap on
finer spatial scales (Hughes~\cite{hughes01}). In the analysis of the moderate
resolution CCD ASCA spectrum of \n, Hughes et al.~(\cite{hughes}) found strong
emission lines from Si, S, Ar, Ca, and Fe, while no emission from O, Ne, or Mg
was required in order to fit the ASCA spectrum. Hughes et al.~(\cite{hughes})
thus concluded that \n\ is the result of a Type Ia supernova. However, its
proximity to the star cluster NGC 1850 and the H II region DEM 84 suggest that
\n\ has an early-type progenitor (Chu \& Kennicutt~\cite{chu}), providing
contradictory evidence for the nature of the progenitor.

In this work, we present the X-ray spectra of \n\ measured by the \xmm\ suite of
scientific instruments, i.e. the Reflection Grating Spectrometers (\rgs) and
the European Photon Imaging Cameras (\epic). Particularly, we exploit the high
dispersion of the \rgs\ and its unique capability to resolve line emission from
extended sources to investigate the nature of the hot plasma in \n\ in
unprecedented detail and to shed new light on its progenitor. The large
effective area of the \epic\ cameras at high energies also allows us to study
the important Fe-K emission. All these are complemented by the sharp images
produced with the high angular resolution of the Advanced CCD Imaging
Spectrometer (\acis) on board \chandra, which reveal the temperature dependent
morphology of \n.

\section{Observation and Data Reduction}

\n\ was observed by \xmm\ on 7 August 2000. The effective exposure times and
filters used on the different instruments are summarised in
table~\ref{tab:tab1}. The \rgs\ data were processed using custom software that
is identical in function to the RGS branch of the Science Analysis System (SAS),
which was described briefly by Rasmussen et al.~(\cite{andy}). It includes
pixel by pixel CCD event offset subtraction using median diagnostic frames,
position dependent CTI correction, and correction of dispersion angles imparted
by aspect drift.

\begin{table}[!ht]
\caption{\xmm\ observation data of \n}
\label{tab:tab1}
\centerline{
\begin{tabular}{|c|c|c|}
\hline
Instrument & Filter & Exposure Time \\
 & & (ks) \\ \hline
MOS 1 & Thick & 8.0 \\
MOS 1 & Medium & 8.0 \\
MOS 2 & Medium & 23.5 \\
PN & Medium & 9.5 \\
RGS 1 & & 32.3 \\
RGS 2 & & 32.4 \\
\hline
\end{tabular}
}
\end{table}

We chose this analysis path since our response matrix generator features support
for extended targets, and details in the angular distribution of \n\ were
included in the response matrices. A single, broad-band archival ROSAT HRI
image was used, along with mean spacecraft aspect and our spectral extraction
angular width, to predict the monochromatic response in the readout channel
space. This modelling approach breaks down as soon as the target's morphology
depends strongly on wavelength. However, it still permits accurate measurement
of total line intensities and significant velocity distributions that distort
line profiles in the \rgs. In the case of \n, wavelength dependent variations
in morphology appear on scales comparable to the \xmm\ mirror point spread
function (PSF), and our modelling approach remains reasonable for the purposes
described. However, we do still expect to observe small ($\sim$20~m\AA)
wavelength shifts for some lines due to variations in the remnant's morphology.

The raw \epic\ data were initially processed with the \xmm\ SAS. This involved
the subtraction of hot, dead, or flickering pixels, as well as removal of events
due to electronic noise. The spectrum was extracted from a circular region with
a radius of 1.5\arcmin. Background subtraction for \epic\ was done using an
exposure of the Lockman hole, with similar data selections and the same
extraction region as for the SNR, scaling according to the exposure time.
Periods of high particle background were rejected based upon the $10-12$~keV
count rate for the entire field of view.

In order to support our spectral analysis of the entire remnant with spatially
resolved information on the morphology, we also used archival \chandra\ data of
\n, which was observed with the back-illuminated ACIS-S3 chip on 4 December 1999
(see also Lewis et al.~\cite{lewisa}). The cleaned event list was taken from
the public \chandra\ archive. Narrow band images were extracted with energy
selections made according to the nominal energy column, which provides a good
energy estimate of each recorded photon.

\section{Data Analysis} 
 
\subsection{\xmm\ Spectra} 
 
\begin{figure} 
\resizebox{\hsize}{!}{\includegraphics[angle=-90]{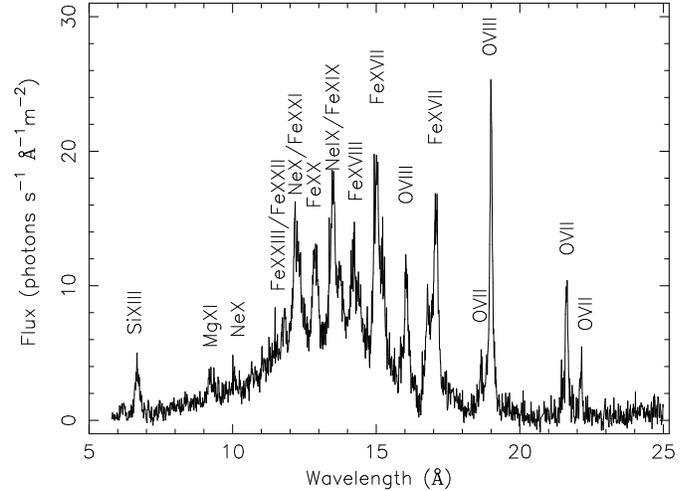}}
\caption{The fluxed \rgs\ of \n\ spectrum. The most prominent lines are 
labeled.} 
\label{fig:fig1} 
\end{figure}

\begin{figure} 
\resizebox{\hsize}{!}{\includegraphics[angle=-90]{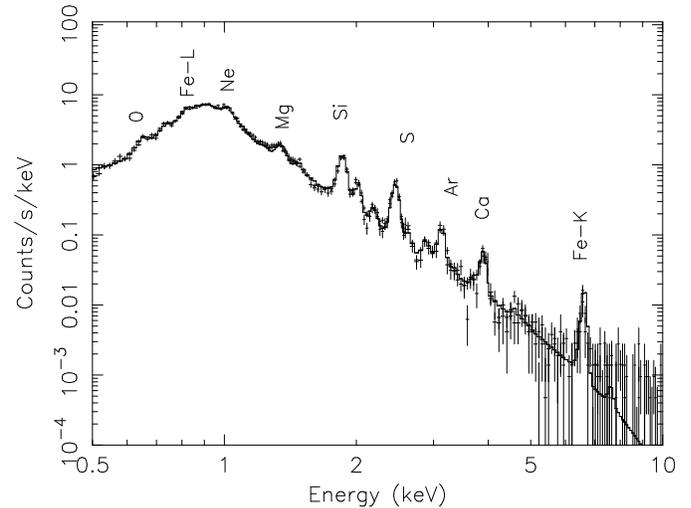}} 
\caption{The \mos\ of \n\ spectrum. The solid line is the best-fit 3 
component NEI model. The most prominent line blends are labeled.} 
\label{fig:fig2} 
\end{figure}

The \rgs\ spectrum of \n\ is shown in figure~\ref{fig:fig1}. The spectrum is
dominated by emission lines from highly ionised species of O, Ne, Mg, Si, \& Fe.
No emission longward of 23~\AA\ is detected. This can probably be attributed to
line-of-sight absorption towards the LMC rather than to the absence of N and C
species, which could have been readily detected by the \rgs. An important point
to note is that we clearly detect O, Ne, and Mg lines. The \ion{O}{vii} and
\ion{O}{viii} lines are particularly strong. These features were not detected
in the CCD spectrum of \n\ obtained with ASCA (Hughes et al~\cite{hughes}), nor
in a more recent CCD measurement with \chandra\ ACIS (Lewis et
al.~\cite{lewisa}). In order to observe shorter wavelengths than provided by
the \rgs\, we also employ the \mos\ CCD spectrum, which is presented in
figure~\ref{fig:fig2}. Emission lines from He-like S, Ar, Ca, as well as an
Fe-K blend at 6.5~keV are clearly observed.

The spectral analysis was performed using the SRON SPEX package (Kaastra et
al.~\cite{kaastra}), which contains the MEKAL atomic database (Mewe et
al.~\cite{mewe}) for thermal emission. Both Coronal Ionisation Equilibrium
(CIE) and non-equilibrium ionisation (NEI) models are available in SPEX. Our
model for the X-ray emission of \n\ consists of several plasma components. For
each component, we fit for the volume emission measure ($n_{\rm e}n_{\rm H}V$),
the electron temperature ($T_{\rm e}$), the elemental abundances, the redshift
of the source, and the Hydrogenic column density $N_{\rm H}$ of neutral
absorbing gas along the line of sight. In NEI models, the ionisation age
($n_{\rm e}t$) (Kaastra \& Jansen~\cite{net}) is also a free parameter. Here,
$n_{\rm e}$ is the electron density, $n_{\rm H}$ is the hydrogen density, $V$ is
the volume of emitting gas and $t$ is the time since the material has been
shocked and heated to its current temperature. A distance of 51 kpc to the LMC
is assumed (Feast et al.~\cite{feast}).

In the following, we describe the procedure used to fit the total spectrum
emitted by the whole remnant. The entire procedure was carried out for both CIE
and NEI scenarios for comparison. We fitted the \rgs\ spectrum first, since it
has the most details and therefore is most constraining. Two plasma components
are sufficient for obtaining a good fit to the \rgs\ spectrum. Assuming that
most of the X-ray emission is due to gas with more or less similar composition,
we coupled the elemental abundances in the two components. This also minimises
the number of free variables in our fit, which makes it more robust.

It is clear from the \epic\ spectrum that \n\ also has a hot component that is
hard for the \rgs\ to distinguish. In order to model this component, we fit the
CCD spectrum of \epic\ with three plasma components. We coupled the elemental
abundances of O, Ne, Mg, Si, S, Ar \& Ca of the third component to those of the
two cooler components. The Fe abundance, however, is left free since we cannot
fit the Fe-L and Fe-K with the same abundance value. The electron temperature,
ionisation age and abundances of the two cooler components are fixed at the
values obtained by the \rgs\ fit. We note, however, that the hot component is
not very sensitive to the other two components even when all three components
are fitted simultaneously. In principle, the \epic\ spectra could be fitted
well with only two plasma components, i.e., without the relatively cold one.
This explains why previous CCD observations did not detect the low-temperature
component, which is dominated by O emission. It is only because of the spectral
resolution of the \rgs\ and its sensitivity to soft X-rays ($>$~20~\AA) that we
are able to resolve the \ion{O}{vii} emission and hence need a cool or
under-ionised component in the model. Finally, for consistency and in order to
obtain the ultimate fit, we return to the \rgs\ spectrum and refit it with three
components. The hot component, which is constrained by the \epic, but hardly
produces line emission in the \rgs\ spectrum is fixed at its \epic\ values. The
final parameters obtained for the three components are presented in Table~2 for
both the CIE and NEI cases. We obtain a best-fit column density of $N_{\rm
H}=(2.5{\pm}0.7){\times}10^{25}$~m$^{-2}$. We also obtain a redshift of
340$\pm$70\ km~s$^{-1}$, which is consistent with the radial velocity of the LMC
($v_{\hbox{LMC}}=278$~km~s$^{-1}$). The significance of NEI as well as the
elemental abundances resulting from the fits are discussed below in \S4.

\begin{table}[!ht] 
\caption{Spectral fitting results. Components 1 \& 2 reflect the \rgs\ 
results, since the \rgs\ is more robust at constraining the parameters in 
this 
spectral range. Component 3 is required to account for the Fe-K emission in 
the 
\epic\ Data.} 
\label{tab:tab2} 
\centerline{ 
\begin{tabular}{|l|c|c|c|} 
\hline 
Parameter		& Covered by	& CIE 	& NEI \\ \hline
Component 1: 	&		&		&	\\
$n_{\rm e}n_{\rm H}V$ (10$^{64}$m$^{-3}$)& & 6.2$\pm$0.8 	& 2.6$\pm$0.1	
\\
$kT_{\rm e}$ (keV) 		&\rgs & 0.21$\pm$0.05	& 0.55$^{+0.05}_{-0.32}$ 
\\	
$n_{\rm e}t$ (10$^{16}$\ m$^{-3}$s)& &		& 2.3$^{+10}_{-0.3}$	\\ 
redshift (kms$^{-1}$) & & 340$\pm$70 & 340$\pm$70 \\ \hline
Component 2: 			& &		&	\\
$n_{\rm e}n_{\rm H}V$ (10$^{64}$m$^{-3}$) &\rgs & 65.1$\pm$1.1	& 65.2$\pm$1.2	
\\
$kT_{\rm e}$ (keV) 	 & + &0.65$\pm$0.05	& 0.65$\pm$0.05	\\
$n_{\rm e}$t (10$^{16}$\ m$^{-3}$s)& \epic\ & 		& $>$250	
\\
redshift (kms$^{-1}$) & & 340$\pm$70 & 340$\pm$70 \\ \hline
Component 3: 			& &		&	\\
$n_{\rm e}n_{\rm H}V$ (10$^{64}$m$^{-3}$) & & 4.1$\pm$0.4& 4.1$\pm$0.5	\\
$kT_{\rm e}$ (keV) 		& \epic\ & 3.5$\pm$0.5 	& 3.5$\pm$0.5 
\\
$n_{\rm e}t$ (10$^{16}$\ m$^{-3}$s)&	&	& 5.3$\pm$0.8 \\
redshift (kms$^{-1}$) & & 4100$\pm$300 & 340$\pm$70 \\ \hline
${\chi}^{2}$/d.o.f & & 1810/902 & 1790/899 \\ \hline
\end{tabular}
} 
\end{table} 
 
\begin{figure*} 
\resizebox{\hsize}{!}{\includegraphics[angle=-90]{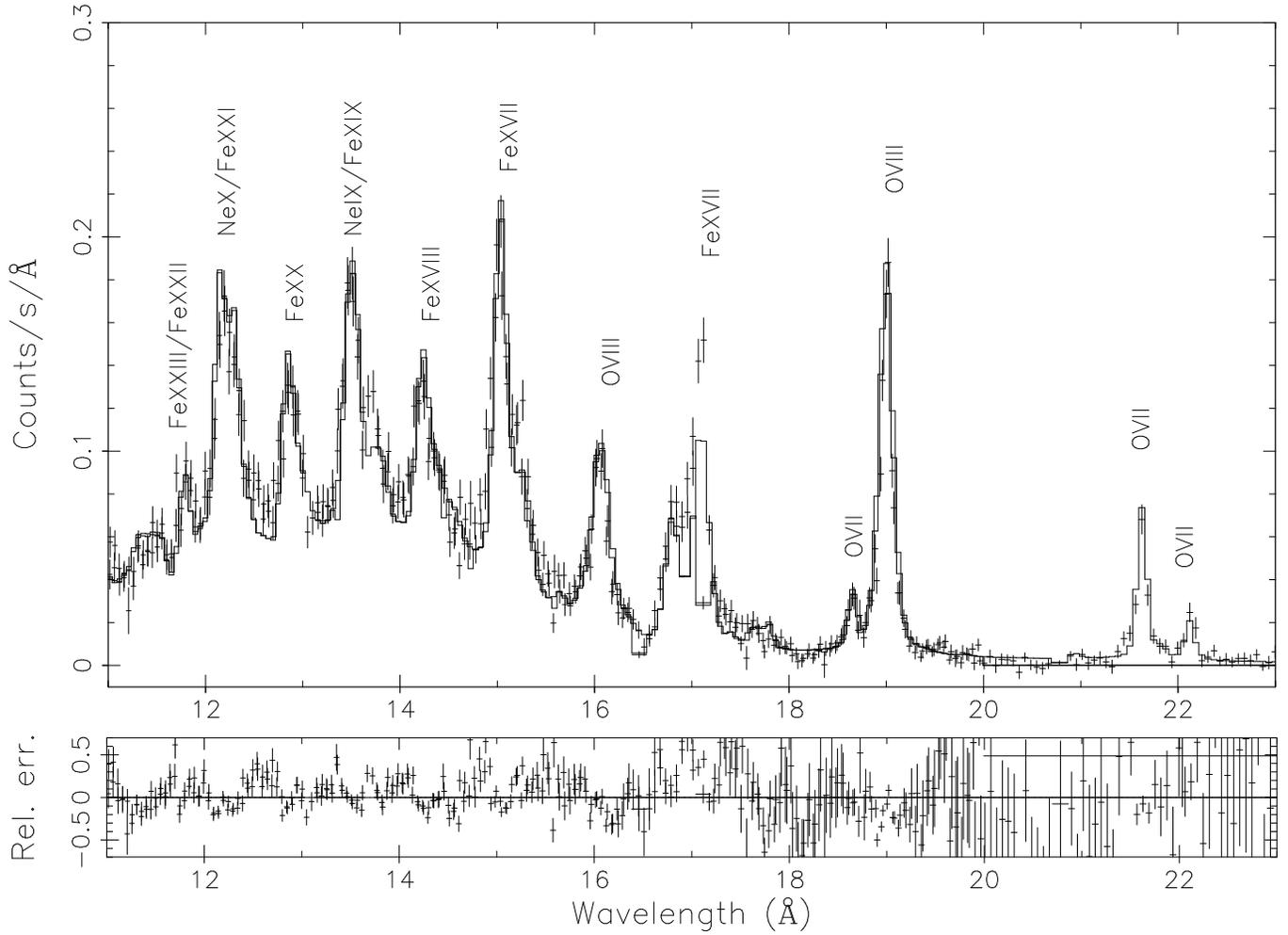}} 
\caption{The \rgs\ spectrum in the range 11-23 {\AA}. The solid line 
represents a best-fit 2 component NEI model ($kT_{\rm e}=$0.65 \& 0.55\ keV) 
plus an underlying $kT_{\rm e}=$3.5\ keV component as obtained from the \mos\ 
fits. The overall fit to the data is good. The missing flux in the model for the 
2p--3s lines of \ion{Fe}{xvii} at $\sim$17\ \AA\ is a known problem which is due 
missing transitions in the plasma code (see text).} 
\label{fig:fig3} 
\end{figure*}

The best-fit spectral NEI model for \n\ is compared with the \mos\ and \rgs\
spectra in figures~\ref {fig:fig2} and \ref {fig:fig3}, respectively. The CIE
model fits the data nearly as well. The overall fit to the data is very good.
For the most part, the fit residuals are consistent within instrumental
calibration uncertainties. However, some small discrepancies remain in the fit.
The most prominent discrepancy is the missing flux in the model for the 2p--3s
lines of \ion{Fe}{xvii} at $\sim$17\ \AA. This is a well known problem and has
been observed in many coronal sources and in the laboratory. Therefore, it
probably has nothing to do with NEI effects, although in principle NEI could
enhance the 17\ \AA\ lines with respect to the 15\ \AA\ lines by means of
inner-shell ionisation of Fe XVI. Doron \& Behar~(\cite{doron}) suggest a model
that could rectify the standard collisional-excitation line powers in MEKAL and
in APEC (Smith et al.~\cite{apec}). They show that including resonant
excitation as well as recombination and ionisation processes (in CIE) from
neighbouring ions in the models considerably enhances the 2p-3s lines and
consequently produces the observed line ratios. Most of the other discrepancies
in figure~\ref{fig:fig3} have local asymmetrical profiles, which are a result of
a combination of the following: (i) difference between the broadband spatial
profile used in the response and the actual spatial profiles of individual
emission lines, (ii) slight errors in model or data wavelengths, (iii) slight
errors in the line spread function which does not account for the real
asymmetrical profile of the RGS line shapes. The first of these sources of
errors (i.e. the spatial profile used) should, however, dominate. There is
also a slight overestimation of the continuum (in the range 18--23\ \AA) in the
models, which is most likely due to uncertainties in the background subtraction.
These inconsistencies between data and models are small and do not affect the
conclusions drawn in this paper.

\subsection{\chandra\ Images} 
 
\begin{figure*} 
\resizebox{\hsize}{!}{\includegraphics{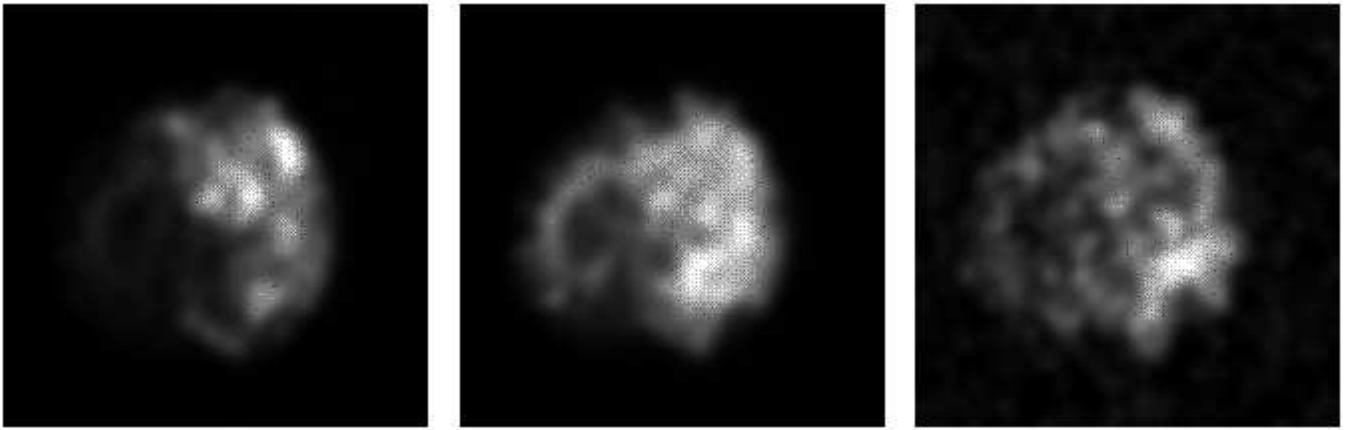}} 
\caption{Chandra images for the approximate energy range 0.5--0.7 keV 
(left), 
0.7--1.0 keV (centre) and 3--6.8 keV (right). These three energy bands are 
dominated by emission from \ion{O}{vii} \&\ ion{O}{viii} (0.5--0.7 keV), Fe-L 
(0.7--1.0 
keV)and 
Ar, Ca, Fe-K and continuum (3--6.8 keV).} 
\label{fig:fig4} 
\end{figure*}

In figure~\ref{fig:fig4} we present three narrow-band images obtained with the
\acis\ camera. The images are produced in the energy ranges of 0.5--0.7,
0.7--1.0, and 3.0--6.8~keV. The first image (0.5--0.7~keV) is essentially an
image of O-K emission (O~VII and O~VIII). The second image (0.7--1.0~keV)
represents predominantly Fe-L emission and the third image (3--6.8~keV) reflects
the morphology of Ar-K, Ca-K, and Fe-K lines, as well as some continuum. We
have also separated this high-energy band into its various sub-component images
(Ar, Ca, and Fe). However, within the statistics available for these images,
all their morphologies appear to be fairly similar. The three images in
figure~\ref{fig:fig4} correspond roughly to the three plasma components
described in Table~\ref{tab:tab1}, in respective order. For short, we will
henceforth refer to these three energy bands (and their respective plasma
components) simply as the O-K, Fe-L, and Fe-K bands (components).

\begin{figure} 
\resizebox{\hsize}{!}{\includegraphics{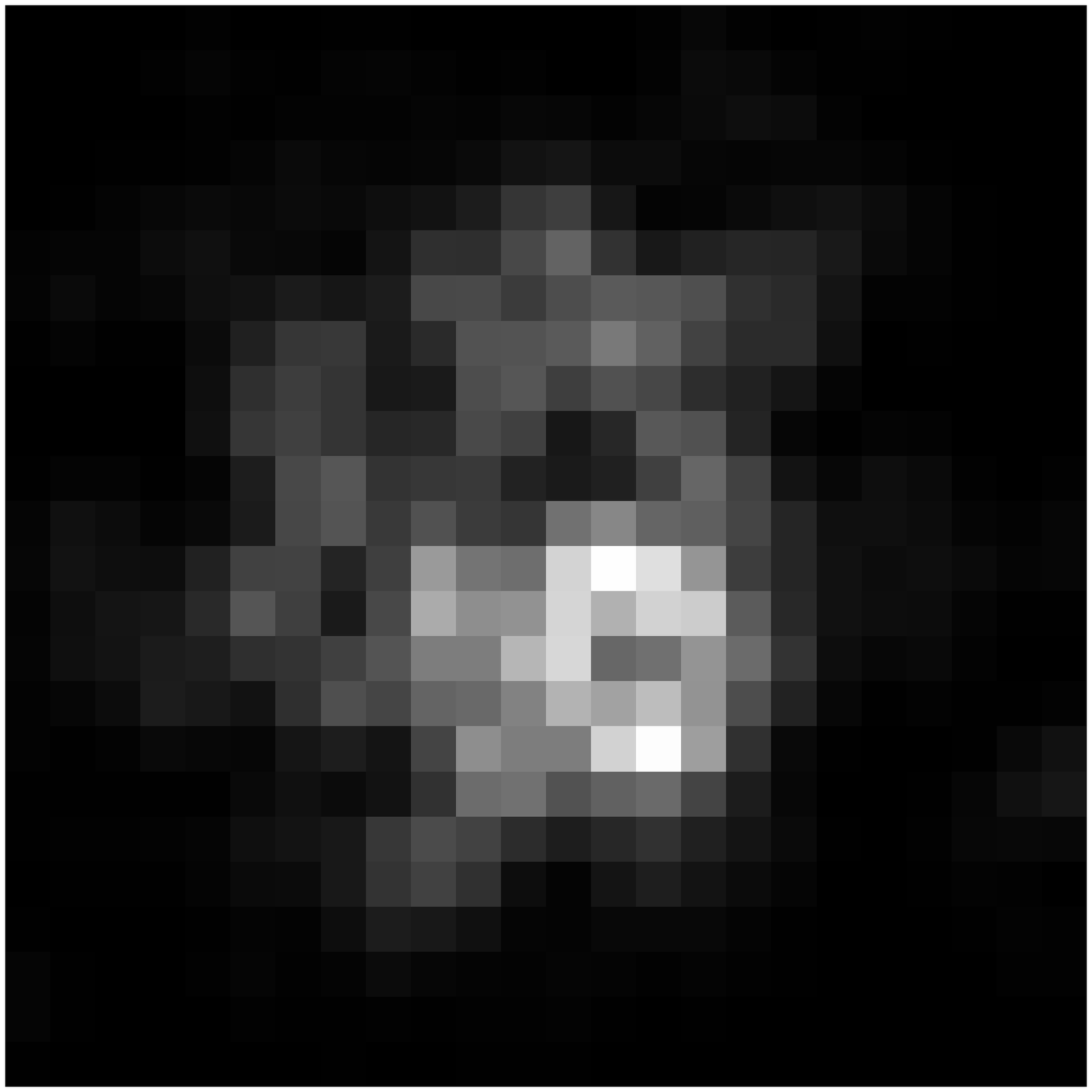}} 
\caption{The deconvolved \xmm\ Fe-K image. The Fe-K image reveals a bright 
emission spot towards the southwestern rim of the remnant.} 
\label{fig:xmmfek} 
\end{figure}

All three images in figure~\ref{fig:fig4} show the same asymmetry, where the
western hemisphere is much brighter than the eastern hemisphere. Aside from
that, they are quite different in morphology. The O-K emission is concentrated
in the northwest, while the Fe-L emission is more evenly distributed along the
western rim of the remnant. The Fe-K image reveals a bright emission spot
located at the southwestern rim. We also created a continuum subtracted,
deconvolved, \xmm\ Fe-K (i.e 6.3--6.8 keV) image (see van der Heyden et al.
\cite{kurt} for details). The \xmm\ Fe-K image (figure~\ref{fig:xmmfek}) also 
shows
a bright emission spot located at the southwestern rim, but extends further
inward than seen in the \chandra\ 3--6.8 keV image. However, while the
statistics are of the \xmm\ image is superior the ability to extract more
information on the morphology of the remnant is hampered by the lower spatial
resolution of the XMM cameras.

\begin{figure} 
\resizebox{\hsize}{!}{\includegraphics{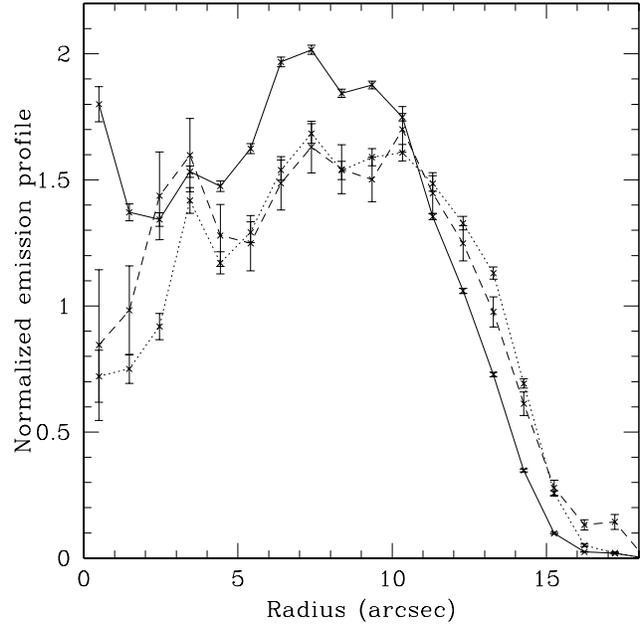}} 
\caption{Normalised radial emission profiles for three separate photon-energy 
bands: 0.5--0.7 (dashed curve), 0.7--1.0 (solid curve), and 3--6.8~keV (dotted 
curve).} 
\label{fig:fig5} 
\end{figure}

A more subtle difference between the emission in the three different bands is
their radial distribution. In order to obtain a further insight into the radial
distribution, we plotted the normalised radial profiles in counts per second in
the three bands. These profiles in counts per pixel are presented in
figure~\ref{fig:fig5}. Each profile is normalised separately to the average
count rate measured in its energy range within a radius of 17.5\arcsec. Despite
the differences in azimuthal distribution, it can be seen that the (normalised)
brightness of all three components is surprisingly similar; peaking at
approximately 4\arcsec, 7\arcsec, and~10\arcsec. The Fe-L component (dashed
curve) lies slightly (0.5\arcsec), but consistently inside the other two
components. Normalising the radial profiles to their peak values does not alter
this conclusion.

\section{Discussion} 
 
\subsection{Ionisation equilibrium \label{sec:equi}} 
 
As demonstrated by the fitting results in Table~\ref{tab:tab2}, the \rgs\
spectrum can be described similarly well by either a set of NEI or CIE plasma
components, giving ${\chi}^{2}$/d.o.f. of 1790/902 and 1810/899, respectively,
for the two methods. Moreover, both models yield very similar spectra. In
terms of emission measure, the X-ray spectrum of \n\ is dominated by the Fe-L
component (2). This emission component is represented by an electron
temperature of $kT_{\rm e}=0.65$\ keV in both models. The NEI model yields a
very high ionisation age ($n_{\rm e}t\ge2.5 \times 10^{18}$~m$^{-3}$s) for this
component, which is sufficiently high for the plasma to be in or very close to
equilibrium (see Mewe~\cite{mewe2}).

\begin{figure} 
\resizebox{\hsize}{!}{\includegraphics[angle=-90]{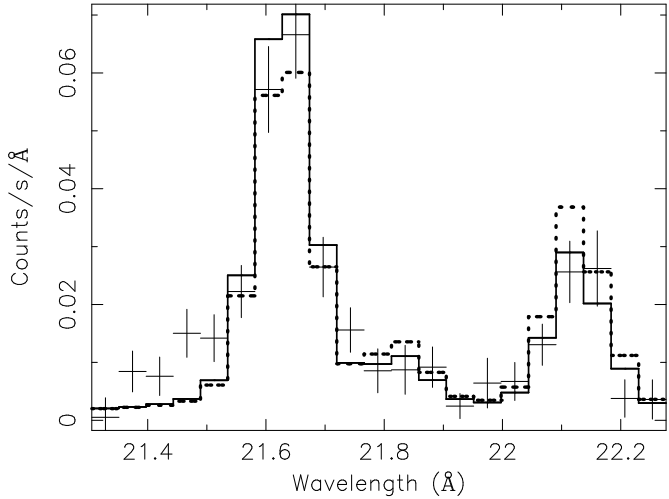}} 
\caption{Comparison of NEI and CIE fits to the \ion{O}{vii} Helium-like triplet. 
The 
crosses are the \rgs\ data points in the 21.3-22.3 {\AA} range. The solid line 
is the $kT_{\rm e}=0.55$\ keV NEI model while the dashed line is the $kT_{\rm 
e}=0.2$\ keV CIE model. Both models provide fits of similar quality, making 
it 
difficult to distinguish between the models.} 
\label{fig:fig6} 
\end{figure}

For the weaker and cooler O-K component (1), which produces the O~VII and most
of the O~VIII emission, the two methods actually give very different temperature
and emission measure values. In CIE conditions, these O ions form at a
relatively low temperature (0.2~keV in the best-fit CIE model). In the case of
NEI, on the other hand, this component is much hotter (0.55~keV), but still
ionising (n$_{\rm e}$t$\sim$2.3$\times$10$^{16}$~m$^{-3}$~s). Both models
reproduce the observed line intensities equally well. Note that the very large
errors associated with the temperature (-0.32~keV) and ionisation time
(+1e17~m$^{3}$s) in the NEI model are due to the fact that the fitting routing
finds another minimum at $\sim$0.2~keV in the case of equilibrium ionisation, as
expected. A comparison of the best-fit models for the O~VII lines in both
approaches is shown in figure~\ref{fig:fig6}. The differences between the NEI
and CIE fits can be seen to be small, which implies that the data do not
decidedly favour one scenario over the other. This CIE/NEI ambiguity in the
O~VII lines was recognised before in the analysis of the \rgs\ spectrum of the
LMC remnant N132D (Behar et al.~\cite{ehud}). It should be noted that the
ambiguity of NEI/CIE in the O-K component (1) does not affect the fits for the
other components at all, i.e., we get the same parameters for components 2 and
3, regardless of whether component 1 is in NEI or CIE.

\begin{figure} 
\resizebox{\hsize}{!}{\includegraphics[angle=-90]{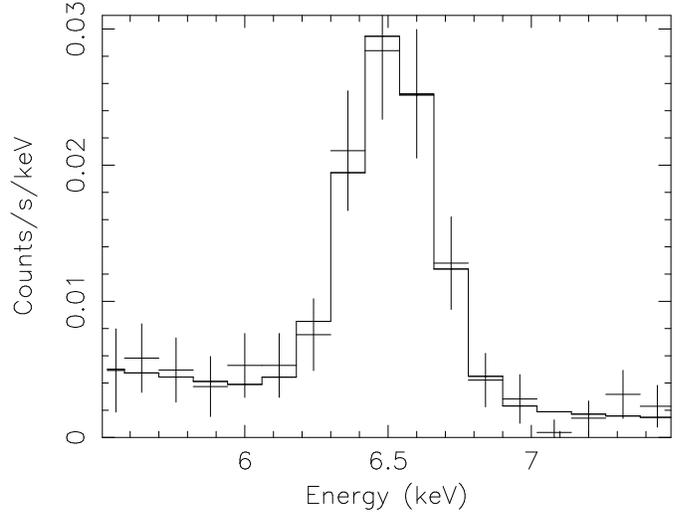}} 
\caption{The \pn\ spectrum in the range 5.5-7.5 keV, showing the Fe-K plot. 
The 
solid line is a $kT_{\rm e}=3.5$\ keV model fit. The location of the line 
centroid ($\sim$6.5\ keV) indicates that the gas is not in ionisation 
equilibrium.} 
\label{fig:fig7} 
\end{figure}

The most outstanding evidence for NEI conditions emerges from the hot Fe-K
component (3), which is responsible for the Fe-K feature measured at 6.5~keV, as
displayed in figure~\ref{fig:fig7}. An attempt to model this feature as a CIE
He-like Fe line (restframe energy 6.7~keV) requires that it be redshifted by
$\sim4100$~km~s$^{-1}$ with respect to the LMC systemic velocity. Such a large
redshift is not observed for the Ar-K nor the Ca-K lines, which arise from the
same component as Fe-K and overlap spatially. This makes the large redshift
interpretation unlikely. An alternative and more plausible explanation for the
spectral position of the Fe-K complex would be a relatively low ionisation age
of $n_{\rm e}t{\sim}5.3{\times}10^{16}$\ m$^{-3}$s for the hot component, as
obtained by our NEI fit (see Table~\ref{tab:tab2}) without any additional
redshift. This NEI component predicts a distribution of Fe charge states that
peaks between \ion{Fe}{xix} and \ion{Fe}{xxiv}. This component also contributes
most of the flux in the Ar and Ca lines and provides good fits to these lines.

\subsection{Density, age, and dynamics \label{sec:density}} 
 
The electron density can be calculated from the emission measure ($n_{\rm
e}n_{\rm H}V$) if we know the volume of the emitting region. For simplicity, we
generally assume the emitting volume to be a thick spherical shell. Based on
the emission maps provided in figure~\ref{fig:fig4} and on the radial profiles
shown in figure~\ref{fig:fig5}, the shell is taken to have a radius of
$R$~=~15\arcsec\ and to be $\Delta R$~=~R/10~=~1.5\arcsec\ thick (Lozinskaya
\cite{lozinskaya}). At a distance of 51~kpc, it gives a volume of
$V=1.7{\times}10^{51}{\times}f$\ m$^{3}$, where $f$ is a volume filling factor.
This is a very crude estimate of the volume, but it is sufficient for our
order-of-magnitude estimates that follow.

For the dominant Fe-L component (2), which seems to cover most of the western
hemisphere (see figure~\ref{fig:fig4}), we assume a half filled shell or
$f$~=~0.5, which gives an electron density of $n_{\rm e} \sim
25{\times}10^{6}$~m$^{-3}$. Using this density and the ionisation age ($n_{\rm
e}t$) from Table~2, we get for the time since the Fe-L gas was shocked:
$t\ge$~3000~yrs. This time is considerably longer than the previously assumed
age of the remnant (1500~yrs), however, the age derived from the ionisation time 
assumes constant density and does not take into account the expansion of the 
shell.

Since the Fe-K component is concentrated in the southwestern region of the
remnant (see figure~\ref{fig:fig4}), we estimate the filling factor to be
$f$~=~0.25. This gives an electron density of $\sim 8{\times}10^{6}$~m$^{-3}$,
which in turn yields an age of $\sim$200~yrs. This is a relatively short time
and is particularly in contrast with the derived age of the Fe-L component. For
component 1, the bulk of the O-K emission is located towards the NW rim of the
remnant. We again assume a filling factor of $f=0.25$. This yields electron
densities of 7.$1{\times}10^{6}$ and $10.6{\times}10^{6}$~m$^{-3}$ for the NEI
and CIE models, respectively. If the NEI model is preferred, its ionisation
time would then also imply a time of about 100~yrs since shock heating. Note
the comparable ages of the O-K and Fe-K components. On the other hand, the
different locations (figure~\ref{fig:fig4}) and vastly different temperatures
of these two components strongly suggest that they may not be dynamically
connected.

Indeed, each of the three spectral components can be identified spatially. As
discussed earlier and as seen in figure~\ref{fig:fig4}, the various spectral
components show substantial variation in morphology. Given our spectral
analysis and due to the fact that roughly comparable densities are obtained for
all of these components, it is most likely that these variations are due to
gradients in temperature and ionisation ages. The different locations and
temperatures of the three components suggest that they are probably unrelated
dynamically. Despite these morphological differences the azimuthal average
emission profiles show a remarkable similarity. Even more surprising is that,
compared to the Fe L component, the coolest (\ion{O}{vii}/\ion{O}{viii}) and the
hottest component peak at a larger radius, extend further out, and seem to be
the most recently shocked components. This would suggest that these components
are associated with material recently shocked by the blast wave, whereas the Fe
L component may be plasma heated by the reverse shock. However, the
overabundance of Ar, Ca and Fe K suggests that, at least the hot component
consists of ejecta material. Note that this confusing situation is corroborated
by Lewis et al.~(\cite{lewisb}), who, based on imaging spectroscopy with
Chandra, show that these elements have radially outward increasing abundance
gradients. Both the abundance gradient and the radial emission profiles of the
three components do not fit the standard description of young supernova
remnants, in which there is an outer, hot, shocked ISM component, and a cooler
shock ejecta component.

\subsection{Elemental abundances and type of progenitor \label{sec:abun}} 
 
As mentioned previously, \n\ was designated to be the result of a Type Ia SN on
the basis of the lack of O, Ne, and Mg emission in the ASCA spectrum (Hughes et
al.~\cite{hughes}). The \rgs\, however, is able to resolve prominent O, Ne, and
Mg emission. Incidentally, the derived abundances are insensitive to the type
of model we use (NEI or CIE). This strongly suggests that \n\ could in fact be
the result of a core collapse SN. To investigate this further, we compare the
abundance yields as obtained from fits to the \rgs\ spectra to those predicted
by theory for Type Ia and Type II nucleosynthesis models. We also compare these
abundances to the average LMC values to verify that the observed X-rays do not
originate from shocked LMC ISM. All LMC abundance values are taken from Hughes
et al.~(\cite{hughes2}) except for those of Ar and Ca, which are from Russell \&
Dopita~(\cite{russell}) as these values are not supplied by Hughes et
al.~(\cite{hughes2}). We use the relative abundances because they are more
consistent among the various models than the absolute abundances (see Hwang et
al.~\cite{hwang}). The calculated abundances for the Type Ia models, the
classical W7 and delayed detonation (WDD2) model, are taken from Nomoto et
al.~(\cite{nomoto}), while the Type II models for 12 and 20~$M_{\odot}$ zero-age
main-sequence progenitors are taken from Woosley \& Weaver~(\cite{woosley}).
The ratio comparisons are given in Table~\ref{tab:tab3}. The Fe-L abundance is
derived from the $kT_{\rm e}=0.65$ keV component, while the Fe-K is from the
$kT_{\rm e}=3.5$ keV component. All abundances are normalised to Si and relative 
to solar ratios (Anders \& Grevese~\cite{anders}).

\begin{table}[!ht] 
\caption{Element abundance ratios as derived from NEI fits to the data 
compared to the LMC abundances and theoretical yield for Type Ia and 
Type II (for 12 \& 20~M$_{\odot}$ progenitors) models. 
The ratios are all relative to solar (Anders \& Grevesse~\cite{anders}) and 
normalised to Si. The abundances for the Type Ia W7 and WDD2 models 
are from Nomoto et al.~(\cite{nomoto}), while the Type II yields are taken 
from 
Woosley \& Weaver~(\cite{woosley}). The LMC values are from Hughes et 
al~(\cite{hughes2}), except for that Ar and Ca which are from Russell \& 
Dopita~(\cite{russell}). } 
\label{tab:tab3} 
\centerline{ 
\begin{tabular}{|l|c|c|c|c|c|c|} 
\hline 
Abun. &Observed&\multicolumn{2}{c|}{Type 1A}&\multicolumn{2}{c|}{Type II} 
& 
LMC\\ \cline{3-6} 
ratio & & W7 &WDD2 & 
12M$_{\odot}$&20M$_{\odot}$ 
&\\ \hline 
O/Si & 0.45$\pm$0.08 & 0.070 & 0.019 &0.17& 1.14 & 0.61 \\ 
Ne/Si & 0.37$\pm$0.07 & 0.005 & 0.001 &0.13 & 1.08& 0.93 \\ 
Mg/Si & 0.25$\pm$0.06 & 0.061 & 0.019 &0.13 & 2.07& 1.03 \\ 
S/Si & 1.91$\pm$0.21 & 1.074 & 1.167 &1.63 & 0.5 & 1.16 \\ 
Ar/Si & 2.56$\pm$0.64 & 0.751 & 0.935 &2.19 & 0.35& 0.54 \\ 
Ca/Si & 2.91$\pm$0.87 & 0.891 & 1.382 &1.75 & 0.42& 0.34 \\ 
FeL/Si & 0.52$\pm$0.14 & 1.88 & 0.866 &0.23 & 0.30& 1.16 \\ 
FeK/Si & 1.24$\pm$0.44 & 1.88 & 0.866 &0.23 & 0.30& 1.16 \\ 
\hline 
\end{tabular} 
} 
\end{table}

\begin{table}[!h]
\caption[]{Estimated mass yields for O, Si and Fe. All values are in solar 
masses.}
\label{tab4}
\centerline{
\begin{tabular}{|l|c|c|c|c|c|}
\hline
Element	&Estimated&\multicolumn{2}{c|}{Type 1A}&\multicolumn{2}{c|}{Type II} \\ 
\cline{3-6}
 & mass & W7 &WDD2 & 12M$_{\odot}$&20M$_{\odot}$ \\ \hline 
O	 & 0.27 & 0.14 & 0.07 &0.22 & 1.48 \\
Si & 0.032& 0.15 & 0.27 &0.099 & 0.095 \\ 
Fe	 & 0.033& 0.70 & 0.67 &0.059 & 0.075 \\ \hline
\end{tabular}
}
\end{table}

The derived abundance ratios measured for \n\ do not compare well with those of
the LMC. While some elements are underabundant by a factor of a few (Ne, Mg,
Fe), others are grossly overabundant (S, Ar, Ca). The absolute abundances (not
shown) are also larger than those of the LMC, indicating that the observed
emission is by and large due to shocked ejecta. Furthermore, the measured
abundance ratios compare much better with the type ~II models than with the type
Ia models. Particularly, our abundance ratios for O/Si, Ne/Si, and Mg/Si are
almost an order of magnitude larger than those predicted for a type Ia
explosion, while the Fe/Si ratio is about a factor of 3 lower than predicted by
the W7 model and slightly lower than predicted by the WDD2 model.

Next, we check if the total Fe and O masses are consistent with a Type Ia SN.
According to our volume estimates for the Fe-L and Fe-K components (2 and 3) in
the previous section, their emission comes from volumes of $V$ = 0.85 and 1.2
${\times}10^{51}$\ m$^{3}$, respectively. Using these volumes along with the
derived EM and Fe abundance, we calculate an Fe mass of 0.02~$M_{\odot}$ (for
$f=0.5$) from Fe-L and 0.013 M$_{\odot}$ from Fe-K, giving a total Fe mass of
0.033 M$_{\odot}$. Similarly, we use the volume estimates made in section
\ref{sec:density}, the emission measure and abundances of components 1 \& 2 and
2 \& 3 to calculate the total O and Si masses respectively. We obtain a total O
mass of 0.27~M$_{\odot}$ and Si mass of 0.032~M$_{\odot}$. In table \ref{tab4}
we compare the estimated O, Si and Fe masses to model predicted yields for type
Ia and type II SNe. The estimated masses are not in exact agreement with any of
the models, but compare better to the type II than type Ia models. In
particular, the Fe mass estimate is only a factor 2-3 smaller than predicted by
the core collapse models, while being more than an order of magnitude smaller
than can be expected for a type Ia SN. The mass increases only with the square
root of the filling factor (or by $\sqrt{V}$), so it is unlikely that our simple
volume estimates could cause the order of magnitude discrepancy between our Fe
mass estimate and that expected from the Type Ia models.

\section{Summary \& conclusions} 
 
The high resolution \xmm\ emission-line spectra of \n\ can be fitted well with
three plasma components. The dominant component (Fe-L) is already in
equilibrium and can be represented by a $kT_{\rm e} = 0.65$\ keV CIE model. The
fact that 0.65~keV gas has equilibrated, together with its volume estimate,
provides an approximate lower limit to the age of \n, which is found to be
3000~yrs, twice as old as previously estimated. A separate component is needed
to explain the O-K emission and here we find difficulty to distinguish between a
relatively cool medium ($kT_{\rm e}=0.2$~keV) in equilibrium or hotter ($kT_{\rm e}=0.55$~keV) ionising
gas. Both scenarios produce essentially the same spectrum for O~VII. The Fe-K
emission can be represented by a hot $kT_{\rm e}=3.5$\ keV NEI component with a
relatively low ionisation age of $n_{\rm e}t{\sim}5.3{\times}10^{16}$\
m$^{-3}$s, implying recent shock heating ($\sim 200$~yrs). The description of 
the X-ray emission from \n\ in terms of this
three-component model is strongly supported by the different morphologies
revealed with the \chandra\ images in three corresponding energy bands. The O-K
component is concentrated in the northwest, the Fe-L component extends along the
western rim of the remnant, and the hottest component (Fe-K) peaks at a bright
knot in the southwest. The O-K and Fe-K components lie, at least partially,
outside of the Fe-L component, which could imply that the Fe-L component was
shocked earlier, whereas the O-K and Fe-K regions were shocked only recently.
However, the different locations and different temperatures of the three
components suggest that they are probably unrelated dynamically.

The \rgs\ spectrum unambiguously reveals emission from species of O, Ne and Mg.
The presence of these lines and the abundance measurements they provide suggest
that \n\ might actually be the result of a type II SN and not a type Ia as
previously thought. Also, the total O mass is higher than predicted by a type
Ia SNe, while the total Fe mass is much lower than would be expected if \n\ was
the result of a type Ia SN.

\begin{acknowledgements} 
The results presented are based on observations obtained with XMM-Newton, 
an ESA science 
mission with instruments and contributions directly funded by 
ESA Member States and the USA. 
JV acknowledges support in the form of the NASA Chandra Postdoctoral 
Fellowship grant nr. PF0-10011, awarded by the Chandra X-ray Center. SRON is 
supported financially by NWO. 
\end{acknowledgements}

\end{document}